\documentclass[10pt]{article}

\usepackage{times}
\usepackage{graphicx}
\usepackage[superscript,biblabel]{cite}
\usepackage{placeins}

\usepackage{lineno}

\title{Estimating the Physical State of a Laboratory Slow Slipping Fault from Seismic Signals}

\author
{Claudia Hulbert$^{1\ast}$, Bertrand Rouet-Leduc$^{1}$,  Christopher X. Ren$^{2}$, \\ Jacques Rivi\`ere$^{3}$, David C. Bolton$^{4}$, Chris Marone $^{4}$, Paul A. Johnson$^{1\ast}$
\\
\normalsize{$^{1}$Los Alamos National Laboratory, Geophysics Group, Los Alamos, New Mexico, USA}\\
\normalsize{$^{2}$Department of Materials Science and Metallurgy, University of Cambridge,}\\ \normalsize{Charles Babbage Road, Cambridge CB3 0FS, UK}\\
\normalsize{$^{3}$Institute of Earth Sciences (ISTerre), Grenoble Alpes University, CNRS,}\\ \normalsize{38000 Grenoble, France}\\
\normalsize{$^{4}$Department of Geosciences, Pennsylvania State University, University Park, Pennsylvania, USA}\\
\\
\normalsize{$^\ast$C. Hulbert (email: chulbert@lanl.gov) and P. Johnson (email: paj@lanl.gov)}
}

\date{}

\begin{document} 


\baselineskip24pt


\maketitle


\begin{abstract}
Over the last two decades, strain and GPS measurements  have shown that slow slip on earthquake faults is a widespread phenomenon. Slow slip is also inferred from correlated small amplitude seismic signals known as nonvolcanic tremor and low frequency earthquakes (LFEs). Slow slip has been reproduced in laboratory and simulation studies, however the fundamental physics of these phenomena and their relationship to dynamic earthquake rupture remains poorly understood. Here we show that, in a laboratory setting, continuous seismic waves are imprinted with fundamental signatures of the fault's physical state. Using machine learning on continuous seismic waves, we can infer several bulk characteristics of the fault (friction, shear displacement, gouge thickness), at any time during the slow slip cycle. This analysis also allows us to infer many properties of the future behavior of the fault, including the time remaining before the next slow slip event. Our work suggests that by applying machine learning approaches to continuous seismic data, new insight into the physics of slow slip could be obtained in Earth. 
\end{abstract}

\pagebreak


There exists a menagerie of earthquake fault slip behaviors \cite{ide2007scaling,ikari2015spectrum}. These range in slip velocity from  slow, where the event can take place over timescales of months to hours with emitted seismic waves corresponding to small amplitude nonvolcanic tremor and low frequency earthquake (LFE) signals (when recorded); to seismogenic stick slip on time scales of seconds to minutes including supershear rupture \cite{rosakis1999cracks,dunham2007conditions}. In general, the balance of different slip behaviors varies with fault zone frictional properties: slow slip dominates primarily in the deeper, ductile portion of a fault and stick slip dominates in the upper, brittle crust \cite{ide2007scaling,scholz2002mechanics}.

Slow slip couples to locked zones on faults in manners not well understood. If this coupling could be characterized, dramatic improvements in earthquake forecasting could follow.  Some faults exhibiting slow slip are known to be periodic such as portions of the subducting slab in Cascadia \cite{articleRogersDragert}, while many are not, for instance in Guerrero, Mexico \cite{Radiguet}. Tremor and Low Frequency Earthquakes that can accompany slow slip appear to come and go, sometimes associated with measurable slip  and sometimes not \cite{shelly2006low,frank2014using}. At times a slow slip event turns into a seismogenic rupture, for instance preceding the Tohoku earthquake \cite{Ito2013,Ruiz1165}, but in general this does not appears to be the case.  

Seismograms are rich in information regarding the seismic source, propagation path and recording site characteristics \cite{Aki1967}.  However, our ability to probe seismic waves for information regarding the precise fault friction, future activity or future slip magnitude is severely limited, with corresponding limitations on seismic hazard analysis and earthquake forecasts.  Is there information contained in seismic signals that we are simply missing and that could inform faulting physics?   For instance, the remarkable discovery of LFE’s and nonvolcanic tremor in 2002 \cite{Obara2002} as data quality improved along with processing abilities, was a surprise. Indeed the spectrum of slip behaviors was not appreciated until the last 15 years or so \cite{ide2007scaling}.  Here, we report on laboratory experiments in which Machine Learning (ML) is used to deduce fault zone friction, shear displacement, slow-slip event time, and the magnitude of future failure events. This work, as well as previous analyzes of similar data for stick-slips \cite{Rouet2017,Rouet2017ArXiv}, suggests a possible avenue for the discovery of previously unidentified signals from earthquake faults.  

With the tremendous progress in fast computing, data storage and marked advances in machine learning, one can analyze vast quantities of data in novel ways, with a minimum of human bias.  In the following, we analyze laboratory experiments applying ML with the goal of developing new approaches that may be useful for understanding and characterizing the physics, timing, and magnitude of slow slip. Specifically, we analyze continuous seismic signals emanating from laboratory fault zones to determine what can be learned regarding fault friction and earthquake timing.  We develop the methods that can be progressively scaled to Earth.  For this purpose, our approach is to rely on transparent machine learning (ML) approaches to analyze simple experimental configurations, that allow us to infer the underlying physics.   

\section*{Laboratory earthquake machine}

Here we analyze data from a double direct, bi-axial shear apparatus, in which  two fault  gouge layers are separated by a central  block driven by a piston at constant velocity (Figure \ref{fig:fig1} (C)).  The fault system is submitted to a fixed normal load (see Materials and Methods and \cite{Marone1998,Johnson2013,Kaproth2013} for details). During a slow slip event, because the gouge material progressively weakens, the force required to keep this velocity constant  decreases. Thus the  stress and friction progressively decrease through the slow slip event. The associated seismic waveforms are also recorded by piezoceramics embedded in the sample assembly. We rely exclusively on the recorded seismic signals to estimate the fault's stress, the fault's displacement, the time remaining before the next failure and the next event's duration and magnitude. Here we analyze two experiments that exhibit different slip characteristics: one slow but resembling stick-slip and another one considerably slower. In the following, we use 'friction' or 'shear stress' interchangeably, as they are proportional at constant normal load.  

\section*{Machine learning determines fault friction and displacement from acoustic emissions}

\begin{figure}[ht!]
\begin{center}
 \includegraphics[width=16cm,trim= 0 0 0 0]{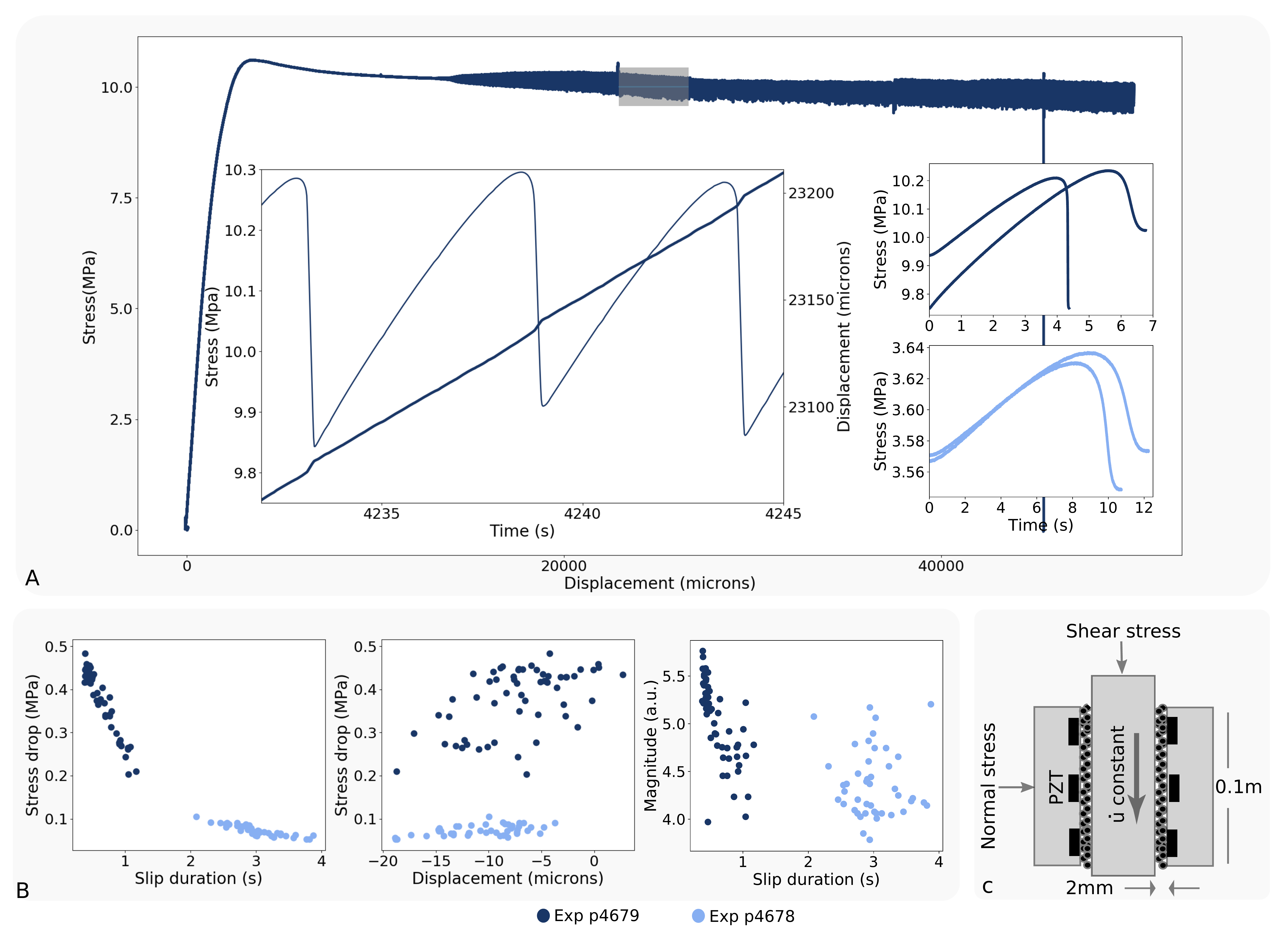}
\caption{\textbf{Slow slip data from a laboratory fault} (\textbf{A})  Upper plot shows shear stress as a function of displacement. The transparent gray box corresponds to the portion of data analyzed, that corresponds to the most aperiodic slow slips. The inset on the left hand-side shows stress versus experimental run time for a few slip cycles (left axis), as well as shear displacement over time (right axis).  At slip time, the shear load decreases and the shearing block advances rapidly forward. Inset on the right shows examples of slip durations observed in each of the two experiments analyzed here. (\textbf{B}) Slip behavior in the experimental data. Stress drops appear to be linearly related to slip duration in both experiments. Weaker relationships seem to link stress drops and fault displacement, as well as maximum amplitude and slip duration. (\textbf{C}) Sketch of the laboratory apparatus.}
\label{fig:fig1}
\end{center}
\end{figure}

Figure \ref{fig:fig1} gives details about the two experiments we analyze. Panel A illustrates the evolution of shear stress in time, and relative displacement. A single experiment is shown as an example in Figure \ref{fig:fig1}(A), along with the shear stress and shear block displacement (left inset). Slip durations range from about 0.1 to nearly 4 seconds for the two experiments. 
Characteristics of the slow slips from both experiments illustrating a range of slip behaviors are shown in Figure \ref{fig:fig1}(B). In both experiments, stress drops appear to be tightly linked to slow slip duration, by a nearly linear relationship (smaller stress drops associated to slower events). There are also connections, albeit weaker, between stress drop and displacement (larger stress drops are associated to more displacement), and the maximum amplitude vs slip durations (smaller amplitudes when slips are slower).

We analyze the continuous seismic signal with a transparent ML approach in order to uncover information related to the underlying physics. By transparent we mean that one can understand exactly how the ML approach makes its estimations.  There are many ML approaches one can apply to such a problem but not all offer clear insight into the underlying physics of the system.  
For this analysis, we rely on a gradient boosted trees \cite{Friedman2000} algorithm, based on decision trees. Decision tree approaches are transparent and can be probed to explain the model estimations (see Materials and Method).

Our goal is to rely exclusively on the continuous seismic signals to infer several key characteristics of the fault (shear stress, shear displacement, time remaining before the next failure, magnitude of the next failure, \textit{etc.}). We apply a moving time window analysis to the continuous seismic data \cite{Rouet2017}. One time window corresponds to $\approx 2.5\%$ of the average duration of one seismic cycle of loading and failure, and successive windows are offset by $0.25\%$ of the average duration of one stress cycle (90\% overlap between two windows). We selected this window length through a grid-search analysis, as it led to the best results. Within one window, we compute many statistical features of the continuous acoustic waveform (thresholds to identify acoustic seismic precursors, higher order moments to measure signal energy, etc. (see Materials and Methods and \cite{Rouet2017} for more details about feature construction, selection, and importance). These features are used as inputs to our ML regression, and are used to determine the fault's instantaneous friction, displacement, or the time remaining before the next failure for a given time window. 
We build our model on the first contiguous half of the data (training set), and evaluate it on the second contiguous half not previously seen by the algorithm (testing set).

During the training phase, the algorithm has access to both the statistical features derived from the seismic signal and the regression label (fault friction, displacement rate or time remaining before failure), and attempts to build a model relating the two (see Materials and Method for more details). As mentioned above, features are exclusively constructed from a small moving window scanning the continuous seismic data. 
The algorithm has access to features from two windows to make one prediction: window N, and window N-1 (previous window with no overlap), providing the means to analyze the short-term evolution of the continuous seismic signal. We first use a statistical analysis (recursive feature elimination \cite{Gregorutti2017}) to select the most promising features by cross-validation on the training set. We then build a gradient boosted trees model from the selected features. The EGO Bayesian optimization method \cite{Jones1998,Rouet2016} enables us to select the hyperparameters of the model (parameters linked to the construction of the ML model, \textit{e.g.} number of decision trees, tree depth,... - see Material and Methods), by 5-fold cross-validation. 

Once the model is built, we evaluate it in the testing phase. Here we use the second half of the data which the algorithm has never seen (testing set). In this phase, the algorithm has access only to the statistical features derived from the seismic signal, and never sees the regression label also measured during the experiment (friction, displacement rate, or time remaining before failure). This label is only used to evaluate the performance of the model estimates. 
We measure the quality of the model by comparing the model estimates to the true experimental friction/stress/displacement values, using the coefficient of determination R$^2$ as an evaluation metric. 

\begin{figure}[ht!]
\begin{center}
\includegraphics[width=16cm,trim= 0 0 0 0]{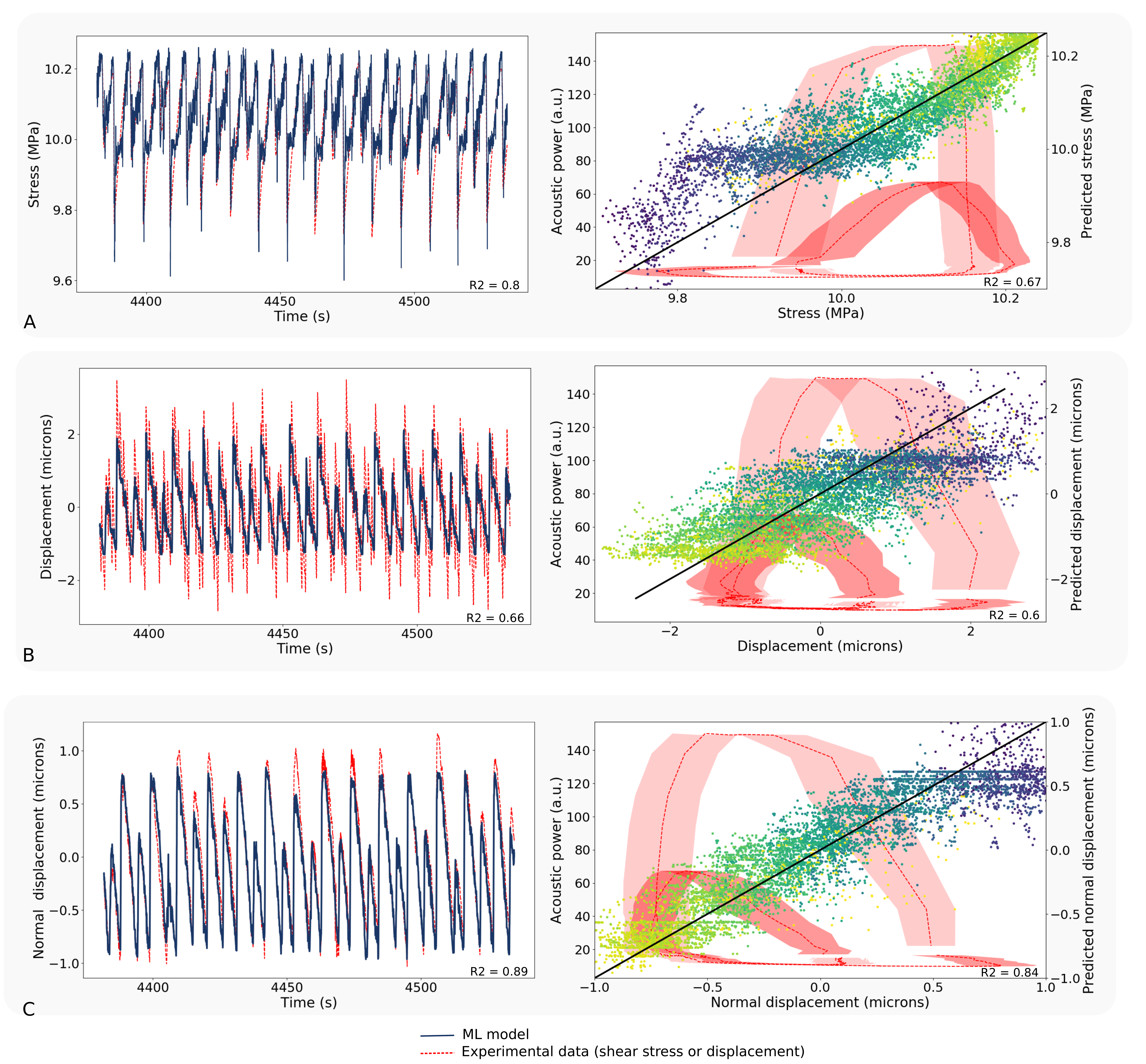}
\caption{Estimation of fault shear stress and detrended displacement (of the shearing and the normal-load blocks) from the continuous seismic emission signal. The normal displacement is a proxy for the thickness of the gouge layer. The left hand-side of panels (\textbf{A}), (\textbf{B}) and (\textbf{C}) show the ML estimates of shear stress (proportional to friction) and fault displacement on the testing set, using all the statistical features extracted from the continuous seismic waves. These estimates are highly accurate, which suggests that the continuous acoustic signal contains a fingerprint of the frictional state of the fault. The right hand-side of panels (\textbf{A}), (\textbf{B}) and (\textbf{C}) show the corresponding estimates for models built exclusively from the best feature identified, the acoustic power. In  shear stress or displacement vs acoustic power space (left axis), all experimental trajectories follow a loop pattern and look similar. Here we plot the average of all stress cycles, and the shades represent a 0.5 standard deviation (not a full standard deviation, for visualization purposes). This shows that the laboratory fault doesn't fail randomly, but follows a very specific pattern leading to failure. The scatter plot (right axis) shows the ML estimates derived from one single feature, the acoustic power. Blue points are farther away from failure, and yellow points are closer to failure. The model estimates improve as failure approaches.}
\label{fig:fig2}
\end{center}
\end{figure}

The model estimates determined (using all the features) for one of the experiments are shown in Figure \ref{fig:fig2}(A),(B),(C) in the left-hand panels (blue curves). The dashed red curves corresponds to the experimental data the algorithm attempts to estimate. Note the bimodal behavior of the slip events  in regard to the magnitude of the stress and displacement where small failures are interlaced with large failures. Bi-modal behavior appears in some slow slip experiments. The ML estimates are shown in blue. Each point of the ML estimation is derived from two short time windows of the continuous seismic signal only, and does not otherwise make use of the signal history nor future. 
By relying exclusively on the instantaneous seismic signal, the ML model is able to accurately estimate the instantaneous shear stress, the displacement of the shearing block, as well as the layer thickness ($R^{2}$ of 0.81, 0.64 and 0.91 respectively) as shown in the left-hand panels. In short, the algorithm  estimates precisely where the system is within the seismic cycle. We emphasize that these estimations are exclusively derived from a portion of the seismic data never before analyzed. 

\FloatBarrier

The ML predictions shown in Figure \ref{fig:fig2}(A),(B),(C) suggest that continuous seismic data are imprinted with precise information regarding the current state of the fault at any time during the laboratory seismic cycle. This holds true for each of the dozens of failure events we analyzed (with different load levels, inter-event times, average slip durations, \textit{etc.}).

Because we rely on an explicit machine learning algorithm, we can probe these models to identify the most important features (see Materials and Methods). For  shear stress, displacement, and layer thickness, the best feature by far corresponds to the variance of the acoustic signal within a time window - \textit{i.e.} the acoustic power. It is straightforward to re-build a machine learning model from this single feature. The three plots on the right hand-side of Figures \ref{fig:fig2}(A),(B),(C) show the performance of ML models that rely only on the acoustic power to make their estimations. The right-hand axis shows the performance of the ML model derived only from acoustic power features (the diagonal black line corresponds to a perfect estimation). The ML algorithm identifies a specific pattern in the acoustic power, that enables it to make these precise estimations. More specifically, the acoustic power of the seismic signal  follows a loop pattern throughout the slow slip cycle. The red dashed lines represent the average loop trajectory in the experimental data for all cycles in the testing set.  The red shaded regions show the 0.5$\times$ standard deviation. The experimental data follow very specific loop patterns in this space reflecting the bimodal behavior of the shear stress (smaller and bigger loops).

By using only the acoustic power, it is still possible to make precise estimations of the current stress and displacement of the fault. Moreover the model estimates improve as failure approaches, as shown by the plots on the right hand-side of Figure \ref{fig:fig2}: here the blue datapoints correspond to times farther away from failure, and yellow to points close to failure.  

These results demonstrate a strong link between the acoustic power emanating from the fault and its frictional state. Because the experimental data follows a very specific pattern for each slip cycle in the friction vs acoustic power space, slow slip events do not occur randomly, but following a specific trajectory that leads to failure.  In short, the fault emits characteristic signals that tell us whether the system is traversing a small or large stress cycle.  

\section*{Machine learning determines upcoming fault failure time and slip duration from acoustic emissions}

\begin{figure}[ht!]
\begin{center}
\includegraphics[width=16cm,trim= 0 0 0 0]{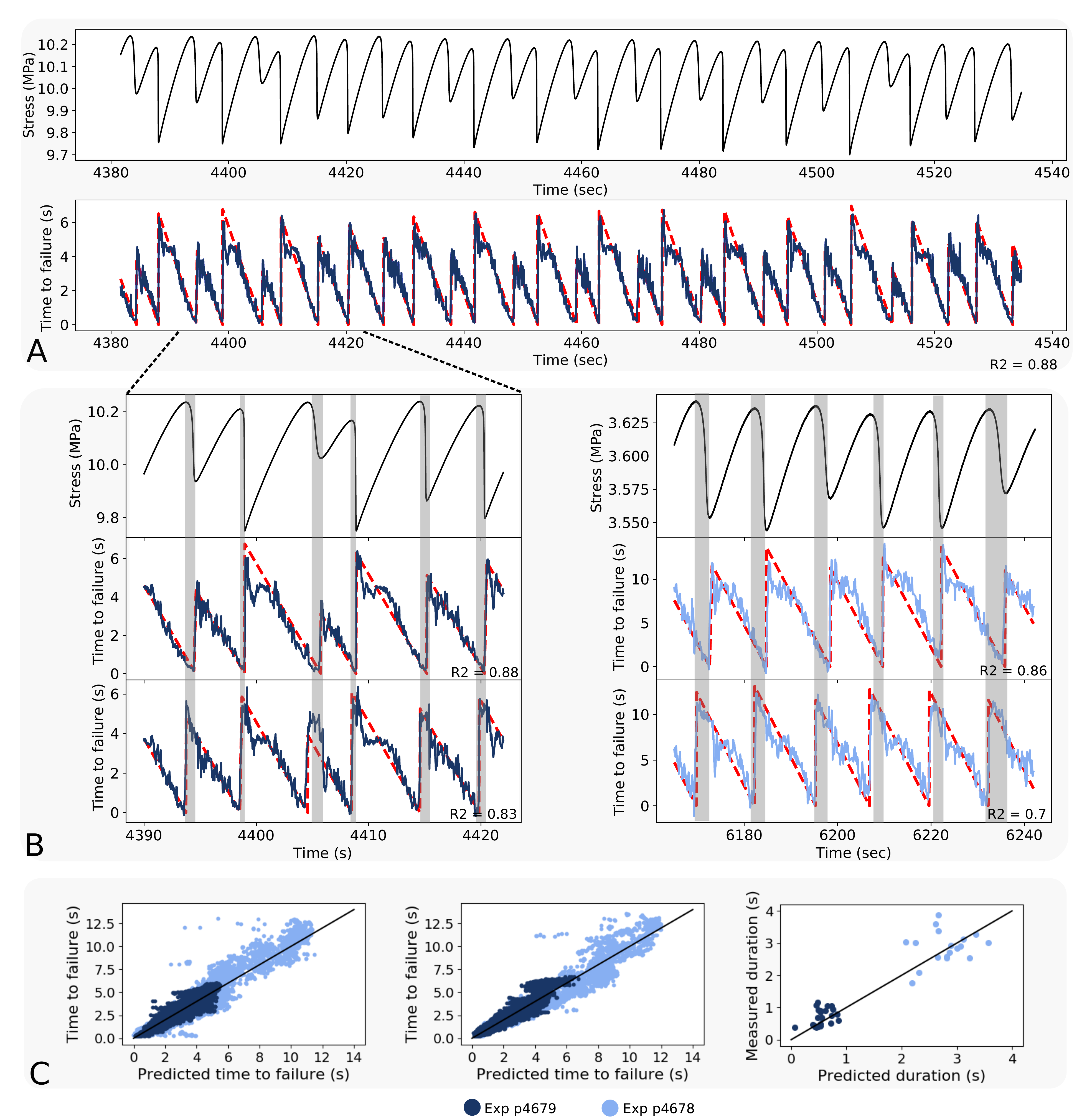}
\caption{(\textbf{A})  Prediction of failure times in a slow slip laboratory experiment. Top panel shows a sequence of slow slips (shear stress in MPa) and bottom panel (blue line) shows the regression analysis of time to failure on the testing set. 
(\textbf{B}) Slip duration derived from predicting the beginning and the end of an event. The two plots correspond to the two experiments. In each plot, the top panel shows the evolution of shear stress with time, the second panel shows the time remaining before the end of the next slow slip event, and the third panel shows the time remaining before the beginning of the next slow slip event. The gray bars delimit the predicted duration of the slip. This ML approach is able to distinguish clearly between longer and shorter events. (\textbf{B}) From left to right: predicted times remaining before the beginning of the next slow slip, predicted times remaining before the end of the next slow slip, and predicted slip duration, versus true experimental values. The black diagonal corresponds to a perfect prediction.}
\label{fig:fig3}
\end{center}
\end{figure}

In our previous work on laboratory stick slip data \cite{Rouet2017}, we found that the time remaining before the next failure can be estimated from the continuous seismic waves emitted from the fault. Here we show that failure timing can also be estimated for laboratory slow slips. The first panel in Figure \ref{fig:fig3} (A) shows the shear stress signal with time for one of the experiments. The bimodal behavior is  clearly seen. The panel below shows predictions of the time remaining before the next failure over the entire testing set for the most aperiodic slow slip experiment. Predictions are highly accurate, with an R$^2$ of 0.88. In particular, the model correctly predicts the time remaining before failure for the two shorter cycles.

When analyzing slow slip, it is possible to predict either the beginning or the end of a slip event. By doing so, one can derive the duration of the slip event from the predictions, as illustrated in Figure \ref{fig:fig3} (B). In this figure, the top panel shows the stress, the middle panel the predictions of the time remaining before the end of the next slip event, and the bottom panel the predictions of the time remaining before the beginning of the next slip event. The gray bars show the duration of the slip, inferred from these predictions. The slip durations inferred from the predictions distinguish long from short slip events. Because the energy of the seismic signals contain quantitative information regarding the stress/friction at all times, the algorithm is able to determine the fault's position within the earthquake cycle, providing the means to infer failure times and event durations long before the slip occurs. Figure \ref{fig:fig3} (C) shows estimated versus true experimental values for time to failure and future event duration.

\FloatBarrier

The accuracy of the failure time predictions, and the fact that we rely on an explicit machine learning algorithm, allows us to gain insight into the  slip physics. More specifically, the features identified by the algorithm as most important for generating these timing predictions are the same as those providing the instantaneous prediction of friction, displacement, and layer thickness (see Figure S2). In particular, the acoustic power and other higher order moments are by far most informative features identified. Because higher order moments are directly linked to the energy of the signal, the fact that the algorithm primarily relies on them for its predictions shows that seismic energy follows a  precise pattern during the stress cycle. Specifically, the acoustic power increases progressively as failure approaches, and decreases as the slip event terminates. By using the acoustic power in the current time window, and in the previous non-overlapping window, the algorithm is able to distinguish between the loading and slipping portions of the cycle. The evolution in time of the strongest features is shown in Materials and Methods. This very specific pattern allows us to make accurate predictions for both the beginning and the end of the slip event.

\section*{Machine learning predicts upcoming fault failure amplitude from its acoustic emissions, predicted inter-event times and predicted event durations}
The accuracy of the failure time predictions enables us to estimate future slip-event amplitudes. We use the predicted inter-event times, the predicted durations of slip events, the variance of the acoustic signal, and the magnitudes of the preceding slip as features for a second machine learning analysis. Thus event amplitude prediction is more difficult than the failure time prediction, as the associated database is much smaller than the database built from scanning the continuous seismic signal (a few dozen datapoints vs several thousands when scanning the seismic signal - see Materials and Methods for more details).

\begin{figure}[ht!]
\begin{center}
\includegraphics[width=16cm,trim= 0 0 0 0]{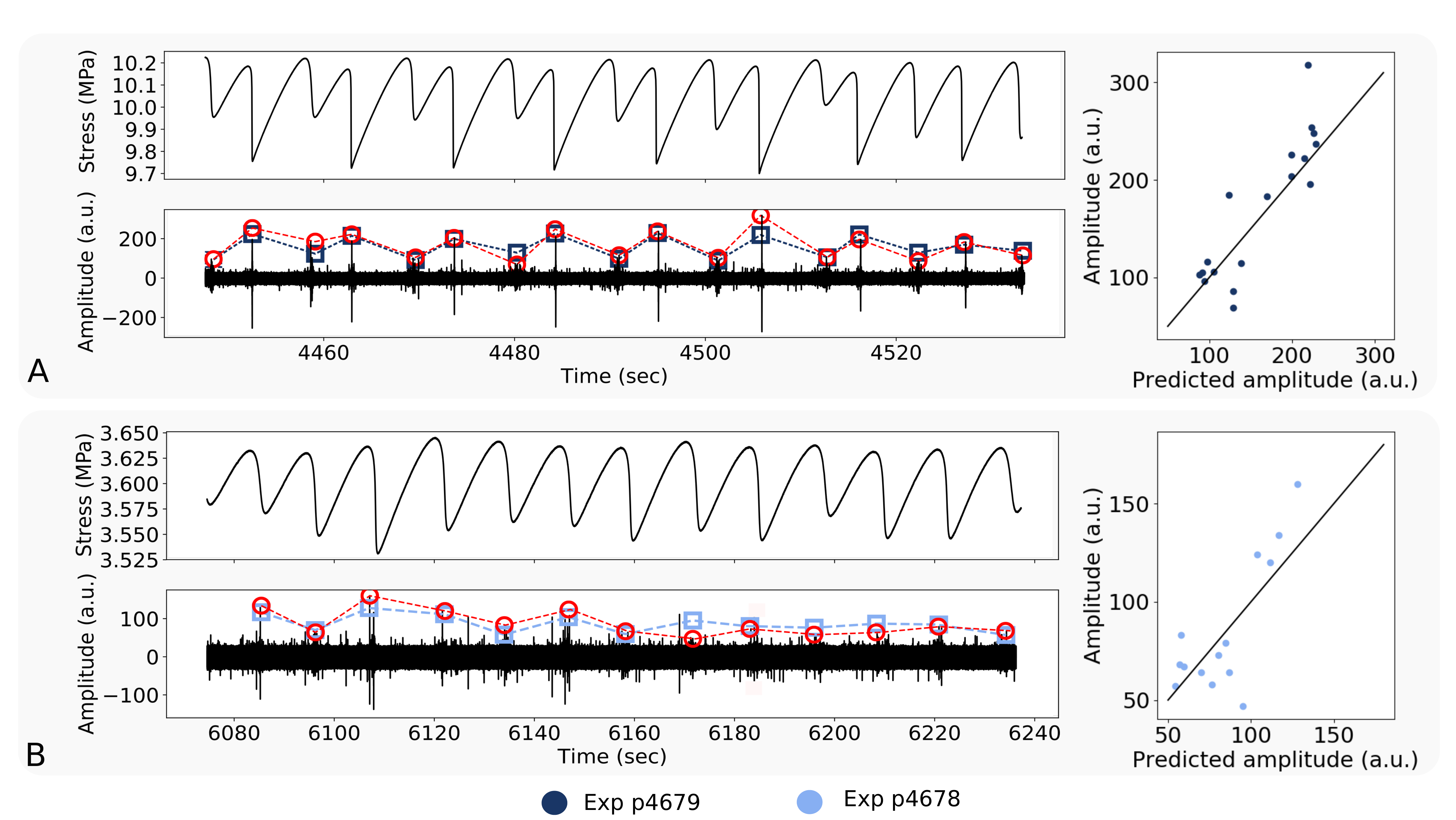}
\caption{Event amplitude predictions on testing set using a second ML analysis. Panels (\textbf{A}) and (\textbf{B}) correspond to the two different experiments.  The top panels show the evolution of the shear stress and the bottom panels show the corresponding seismic signal.  The red circles correspond to the actual amplitudes and the blue squares show the predicted values. Note that as we predict the amplitude of the absolute value of the signal, the highest amplitude may correspond to the negative portion of the signal. Plots on the right hand-side show the predictions versus true experimental values.}
\label{fig:fig4}
\end{center}
\end{figure}

Specifically, we predict the maximum half-peak amplitude of the absolute value of the acoustic signal for the next event, $A = \textrm{max}(\textrm{abs}(ac))$, which can be used in turn to calculate magnitude. We rely on another gradient boosted trees regression to make the predictions, as this approach resulted  again in the highest performance when compared to others tested. Figure \ref{fig:fig4} shows predictions on the testing set for the two slow slip experiments analyzed. The open red circles correspond to the true experimental amplitudes, and the blue open squares show the predicted values. Event magnitude $M$ can be obtained from $M=\log(A)$, with $A$ the maximum half peak amplitude. Because we are measuring the signal directly adjacent to the fault zone there is no need to account for the distance from the source.

Although magnitude predictions are not as accurate as the timing predictions due to the much smaller amount of data available, they are nonetheless good. We constructed 50 different models for both experiments; the average R$^{2}$ for the first experiment was 0.73 (for about 50 datapoints), and for the second experiment 0.42 (for about 30 datapoints). The predicted slip inter-event time (as one may expect) and the predicted slip duration appear to be the most important variables to make these predictions. When the duration of the slip is longer, the energy is released over a longer period of time, which may lead to lower acoustic amplitudes.

\FloatBarrier

\section*{Conclusion}

By applying machine learning to data from slow slip experiments, we show that at any time during the slow slip cycle, continuous seismic signals are imprinted with fundamental physical information regarding the current state of the fault. 

Because the important features identified by the ML approach are tightly linked to the signal energy, these results show that the energy of the seismic signal follows a very specific pattern throughout the slow slip cycle. More specifically, the analysis suggests that the signal energy increases imperceptibly early in the stress cycle, accelerates before a slow slip event, and decreases sharply when the end of the event is near. The machine learning analysis identifies this systematic change in energy over time, enabling predictions.

This steady increase in signal energy is likely due to grain-to-grain and grain-to-fault block interactions that broadcast elastic waves from the gouge layers, as in shown in Discrete Element simulations of stick slip \cite{ferdowsi2014discrete,DorostkarGRL,GRL:GRL56055}. It is reasonable to think that slowly slipping faults produce similar noise throughout the slip cycle. If this is so, the question is whether such a signal can be measured at the surface of the Earth.  Measurements of nonvolcanic tremor and low frequency earthquakes suggest that the signal is there to be identified. We are currently investigating this issue with  data from episodic slip in Cascadia. The fundamental question that follows is how to relate coupling of slow slip to the seismogenic zone on locked faults updip, as in Cascadia, which is our ultimate goal.

\section*{Acknowledgements}
We thank institutional support (LDRD) at Los Alamos and  DOE Fossil Energy for funding this work. We thank Joan Gomberg, Andrew Delorey and Robert Guyer for discussions and comments.  

\section*{Author contributions statement}
C.H., B.R.L., and C.R. conducted the machine learning analysis. J.R., C.B., P.J. and C.M. conducted the experiments. P.J. supervised the project. All authors reviewed the manuscript. 

\bibliographystyle{bib_style}


\end{document}